# Humid evolution of haze in the atmosphere of super-Earths in Habitable Zone

Julien MAILLARD[1,2], Nathalie CARRASCO[1*], Christopher P. RÜGER[2], Audrey CHATAIN[1], Isabelle SCHMITZ-AFONSO[2], Chad R. WEISBROD[4], Laetitia BAILLY[2], Emilie PETIT[2], Thomas GAUTIER[1], Amy M. MCKENNA[4] and Carlos AFONSO[2]

[1] Université Paris-Saclay, UVSQ, CNRS, LATMOS, 78280, Guyancourt, France

[2] Normandie Univ, COBRA UMR 6014 et FR 3038 Univ Rouen; INSA Rouen; CNRS IRCOF, 1 Rue Tesnière, 76821 Mont-Saint-Aignan Cedex France

[4] National High Magnetic Field Laboratory, Florida State University, 1800 East Paul Dirac Drive, Tallahassee, Florida 32310, United States

**\*Corresponding author:**

Nathalie Carrasco

Université Paris-Saclay, UVSQ, CNRS, LATMOS, 78280, Guyancourt, France

Tel +33 (0)2 35 52 29 19

Nathalie.carrasco@latmos.ipsl.fr

**Abstract**

Photochemical hazes are expected to form and significantly contribute to the chemical and radiative balance of exoplanets with relatively moderate temperatures, possibly in the habitable zone (HZ) of their host star. In the presence of humidity, haze particles might thus serve as cloud condensation nuclei and trigger the formation of water dropplets. In the present work, we are interested in the chemical impact of such a close interaction between photochemical hazes and humidity on the organic content composing the hazes and on the capacity to generate organic molecules with high prebiotic potential. For this purpose, we explore experimentally the sweet spot by combining N-dominated super-Earth exoplanets in agreement with Titan's rich organic photochemistry and humid conditions expected for exoplanets in habitable zones. A logarithmic increase with time is observed for the relative abundance of oxygenated species, with O-containing molecules dominating after 1 month only. The rapidity of the process suggests that the humid evolution of N-rich organic haze provides an efficient source of molecules with high prebiotic potential.

# 1. Introduction

Aerosols and clouds, photochemical hazes, or dust lifted from planetary surfaces seem to be common, if not ubiquitous, in exoplanetary atmospheres (Gao *et al.* 2021). After first theoretical predictions (Guillot *et al.* 1996; Saumon *et al.* 1996), they are now directly observed in spectral signatures of exoplanetary atmospheres obtained with transmission spectroscopy on hot Jupiters (Charbonneau *et al.* 2002) and with direct imaging on young giant exoplanets (Marois *et al.* 2008). Among aerosols, photochemical hazes are produced by atmospheric chemistry and are expected to be significant atmospheric contributors on planets with relatively moderate temperatures, up to several hundred Kelvin according to laboratory studies (He *et al.* 2018; Hörst *et al.* 2018).

Hazes can directly impact thermal structures of planets, as illustrated on Titan in the solar system where the anti-greenhouse effect of the haze at high altitude reduces the surface temperature of Titan by 9 K (McKay et al. 1991). They can also serve as cloud condensation nuclei and indirectly impact the atmospheric thermal balance through the formation of clouds (Lohmann and Feichter 2005). The wettability property has been experimentally determined for a series of laboratory analogues of hazes representative of expected temperate exoplanetary atmospheres (Yu *et al.* 2021). Some of the haze analogues demonstrated high surface energy with water ($H_2O$), which is consistent with a high propensity for hazes in the atmospheres of planets in the habitable zone to trigger water condensation on their surfaces.

In the present work, we consider the chemical evolution of hazes in the atmospheres of Super-Earths or Earth-like exoplanets in the habitable zone. Dry organic photochemical hazes are expected to contribute to the formation of prebiotic molecules (Moran *et al.* 2020). Their further chemical evolution under humid conditions can directly affect planetary habitability. However, deciphering the chemical composition of atmospheres of Super-Earths and Earth-like rocky exoplanets is a major challenge for incoming James Webb Space Telescope observations. To prepare for these observations, exploratory modeling is proposed that is reliably based on our knowledge of the solar system. In this frame, Titan, the largest satellite of Saturn, offers an extreme case of haze formation and the rocky Super-Earth 55 Cancri e is, for example, suspected to be dominated by molecular nitrogen ($N_2$), like Titan (iiguel 2019). We therefore propose a comprehensive experimental dataset to address the chemical evolution of exoplanetary haze analogues representative of $N_2$-rich atmospheres in humid atmospheric conditions.

The presence of such hazes is also strongly suspected to have occurred during the Archean eon on the early Earth (Arney *et al.* 2016; Haqq-Misra *et al.* 2008; Izon *et al.* 2015; Kasting *et al.* 1989; Kurzweil *et al.* 2013; Trainer *et al.* 2004; Trainer *et al.* 2006; Zerkle *et al.* 2012). Today, the most accepted idea is that the original gas mixture that produced these organic aerosols by photochemistry was composed of $N_2$, $CO_2$, $H_2O$, CO, and a small amount of reducing species including methane ($CH_4$) (Cleaves *et al.* 2008; Kasting 1993; Trainer *et al.* 2004). In specific locations such as volcanic plumes, the primitive atmosphere could have been much more reducing, containing molecular hydrogen ($H_2$) and $CH_4$ in addition to $N_2$ and other gases (Johnson *et al.* 2008; Parker *et al.* 2011; Tian *et al.* 2005). Studying the interaction of such a haze with atmospheric humidity would, therefore, facilitate a better understanding of the prebiotic chemistry that occurred before and during the emergence of life on Earth.

## 2. Experimental methods

To study the evolution process, haze analogues were initially synthesized in dry and oxygen-free conditions. These conditions were chosen to be representative of the model case provided by Titan's atmosphere, which has been relatively well studied in the past. It also serves as a proxy for aerosols produced in dry conditions at high altitude, above the $H_2O$ cold-trap, in the atmospheres of the early Earth or of other worlds in the habitable zone. However, the haze can also directly form in humid conditions at lower altitudes. In this frame, our approach works to decouple exposure to humidity from the incorporation of oxygen occurs during the synthesis step. This enables us to isolate the process of exposure to humidity and parameterize it for further implementation in models.

Five samples were synthesized with the intent to study chemical evolution at five different aging times. Each sample was analyzed by using a 21 Tesla Fourier transform ion cyclotron resonance mass spectrometer (FT-ICR MS). This type of mass spectrometer is state-of-the-art technology is the most suitable for retrieving information on highly complex organic mixtures (Marshall and Chen 2015; Marshall and Rodgers 2008; McKenna *et al.* 2013a; McKenna *et al.* 2013b; McKenna *et al.* 2010; Podgorski *et al.* 2013). In addition, to validate and further our understanding of the chemical functions present in these samples, infrared spectroscopy was performed.

### 2.1. Synthesis and humid aging of exoplanetary haze analogues

Analogues of exoplanetary hazes are produced without water under conditions that simulate a temperate atmospheric photochemical synthesis based on $CH_4$ and $N_2$, which is in agreement with Titan's atmospheric composition. Haze analogues are produced with the PAMPRE plasma experiment following a well-established protocol detailed in previous publications (Szopa et al. 2006). The reactor is composed of a stainless-steel cylindrical reactor within which a RF-Capacitively Coupled Plasma discharge is triggered by an RF 13.56 MHz frequency generator. A gas mixture of 95% $N_2$ and 5 % $CH_4$ is injected into the chamber as a continuous flow through polarized electrodes, and it is further extracted by a primary vacuum pump to ensure that gases are homogeneously distributed. The plasma discharge is maintained at a pressure of $0.9 \pm 0.1$ mbar and at room temperature ($21.0 \pm 0.1$°C). Haze analogues are recovered as a brown powder after 1 day, and stored in glass vials.

After synthesis, the samples are stored in an enclosed dark place. The humid aging process is performed by storing the vials in a box at 20°C, where hygrometry is kept stable at 70%. The box is protected from light and is additionally covered with an alumina sheet to enhance UV protection. The aging is performed by exposing the samples to atmospheric humidity at a hygrometry of 70%. A contribution of oxidation with molecular oxygen ($O_2$) is possible, but it has been shown to be negligible compared **to** the reaction with ambient humidity (see discussion). The samples are aged for periods that range from 1 month to 3.5 years in the absence of light. Five aging time-durations have been defined as follows: 1 month, 4.5 months, 1.8 years, 2.2 years, and 3.5 years after production.

**2.2. FT-ICR Mass Spectrometry**

The analyses are performed on a Fourier transform ion cyclotron resonance mass spectrometer equipped with a 21 Tesla superconducting magnet (Hendrickson *et al.* 2015; Smith *et al.* 2018).

Ions are generated at atmospheric pressure via micro electrospray (Emmett *et al.* 1998). This ionization mode is required to recover a fraction of the sample extracted in methanol. For this purpose, 4 mg of the haze analogue are dissolved in 1 mL of methanol ($CH_3CH_2OH$) in a vial. The vial is vigorously stirred for 3 minutes to solubilize the methanol-soluble species. The brown mixture is then filtered by a 0.2 μm polytetrafluoroethylene (PTFE) membrane filter on a filter holder. The filtered solution is transferred into a vial, then dried with nitrogen and stored until analysis. Samples are dissolved in methanol to produce a stock solution of 500 µg/mL. All experiments are recorded in electrospray (ESI) positive ionization mode.

Ions are initially accumulated in an external multipole ion guide (21 T: 1-5 ms) and released *m/z*-dependently by a decrease of an auxiliary radio frequency potential between the multipole rods and the end-cap electrode (Kaiser *et al.* 2014). Ions are excited to *m/z*-dependent radius to maximize the dynamic range and number of observed mass spectral peaks (*m/z* 150-1500; 21 T: 32-64%) (Kaiser *et al.* 2013), and excitation and detection are performed on the same pair of electrodes (Chen *et al.* 2014). The dynamically harmonized ICR cell is operated with a 6 V trapping potential (Boldin and Nikolaev 2011; Kaiser *et al.* 2011). Time-domain transients of 3.1 seconds are acquired with the Predator data station, with 100-500 time-domain acquisitions averaged for all experiments (Blakney *et al.* 2011). Mass spectra are internally calibrated with a high-abundance homologous series based on the "walking" calibration method (Savory *et al.* 2011).

The need for ultrahigh resolving power only achievable by FT-ICR MS is explained by the chemical complexity of the sample. **Figure 1** shows a mass spectrum in positive ion ESI for the methanol-soluble fraction of the initial sample before aging. Analytes are distributed across a wide *m/z* range (from 180 < *m/z* < 1200) in a log-normal distribution with more than 30,000 peaks detected with signal magnitude greater than six times the baseline RMS noise level, and the repetition patterns observed previously are recovered (Pernot *et al.* 2010; Somogyi *et al.* 2005). In **Figure 1**, a mass-scale zoom insert between *m/z* 558.24 and *m/z* 558.40 highlights the immense compositional complexity, with more than 100 peaks detected within a 0.16 Da mass window. A second insert displays the data over a 0.04 Da *m/z* range with several fully resolved mass spectral peaks that correspond to compounds that differ in mass by roughly the

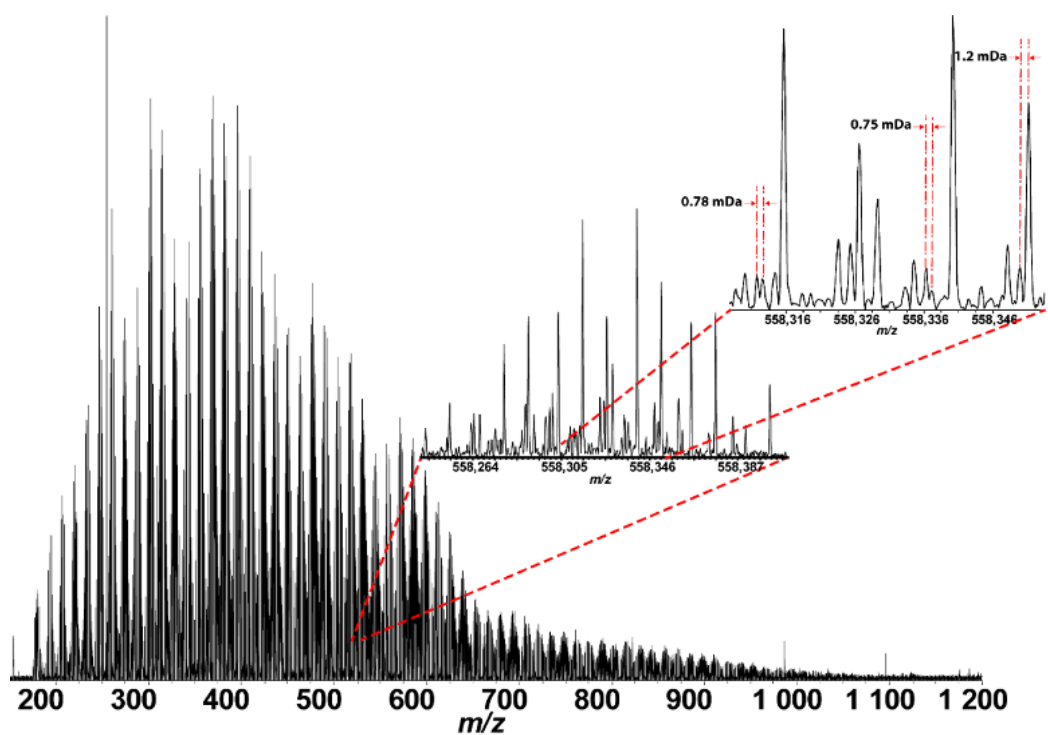

**Fig. 1:** Mass spectrum of the initial sample extracted in methanol and recorded with a 21 T FT-ICR MS.

## 2.3. FT-ICR Data Analysis

Experimentally measured masses are converted from the International Union of Pure and Applied Chemistry (IUPAC) mass scale to the Kendrick mass scale (Kendrick 1963) for rapid identification of homologous series for each heteroatom class (i.e., species with the same $C_cH_hN_nO_oS_s$ content, differing only by the degree of alkylation) (Hughey *et al.* 2001). For each elemental composition, $C_cH_hN_nO_oS_s$, the heteroatom class, the double bond equivalent DBE (number of rings plus double bonds to carbon) (McLafferty and Turecek 1993), and carbon number, *c*, are tabulated for subsequent generation of heteroatom class relative abundance distributions and graphical relative-abundance weighted DBE versus carbon number images. Peaks with signal magnitude greater than 6 times the baseline root-mean-square (rms) noise at *m/z* 500 are exported to peak lists, and molecular formula assignments and data visualization are performed with PetroOrg © software (Corilo 2014). Molecular formula assignments with an error >0.5 parts-per-million are discarded, and only chemical classes with a combined relative abundance of ≥0.15% of the total are considered.

**2.4. Infrared transmission spectroscopy**

Infrared transmission spectroscopy is used to obtain information on the chemical composition of the sample (Chatain *et al.* 2020). Samples are conditioned for IR measurements as a thin deposit on a grid. A total of 1 mg of haze analogues is spread on a thin stainless-steel grid of 1.3 cm in diameter. The grid is composed of threads of 25 μm-large and meshes of 38 μm, that lead to a transmission of 20% in IR wavelengths. The samples are placed inside a glass compartment under vacuum with $CaF_2$ windows, which is inserted into the sample compartment of a Nicolet 6700 FTIR from Thermo. The glass compartment and the sample are then degassed under vacuum during 30 min to remove water traces adsorbed on the sample. The spectra are acquired by a DTGS detector between 900 and 4000 $cm^{-1}$, this range being limited at 900 $cm^{-1}$ by the $CaF_2$ windows used on the FTIR and the reactor, respectively. We chose a spectral resolution of 2 $cm^{-1}$ and an average of 250 scans. The width of the IR beam at the focal point is a few millimetres. The spectra are corrected for $H_2O$ and $CO_2$ atmospheric contributions with the OMNIC software. Baseline is corrected with an adjusted 2$^{nd}$ order polynomial function using two reference wavelength ranges where the samples do not absorb: 1820-1880 $cm^{-1}$ and 3790-4000 $cm^{-1}$.

**2.5. Elemental analysis**

An elemental composition analysis of the bulk sample is carried out with a FLASH 2000 instrument from Thermo to measure the C, H, N, and O mass percentages in the sample under dark aging through time. The instrument is supplied by using a double reactor configuration arranged for simultaneous CHN and oxygen analysis. Oxygen determination is achieved with an oxygen-specific pyrolysis reactor heated at a temperature slightly above 1060 °C. For C, H, and N, the sample is burnt in the presence of oxygen at 1800 °C for a few seconds. At this high temperature, organic and inorganic substances are converted into elemental gases, which, after further reduction, are separated in a chromatographic column and finally detected by a highly sensitive thermal conductivity detector (TCD).

# 3. Results

**Figure 2** shows a comparison between 3 samples from an aging time of 1 month (**Fig 2a**), 1.8 years (**Fig 2b**), and 3.5 years (**Fig 2c**). For all three samples, the molecular weight distribution is centred at *m/z* 558. An increase in the number of peaks over time corresponds to an increase in aging. From 1 month to 3.5 years, a more than two-fold increase in complexity occurs, which indicates the chemical transformation with exposure time.

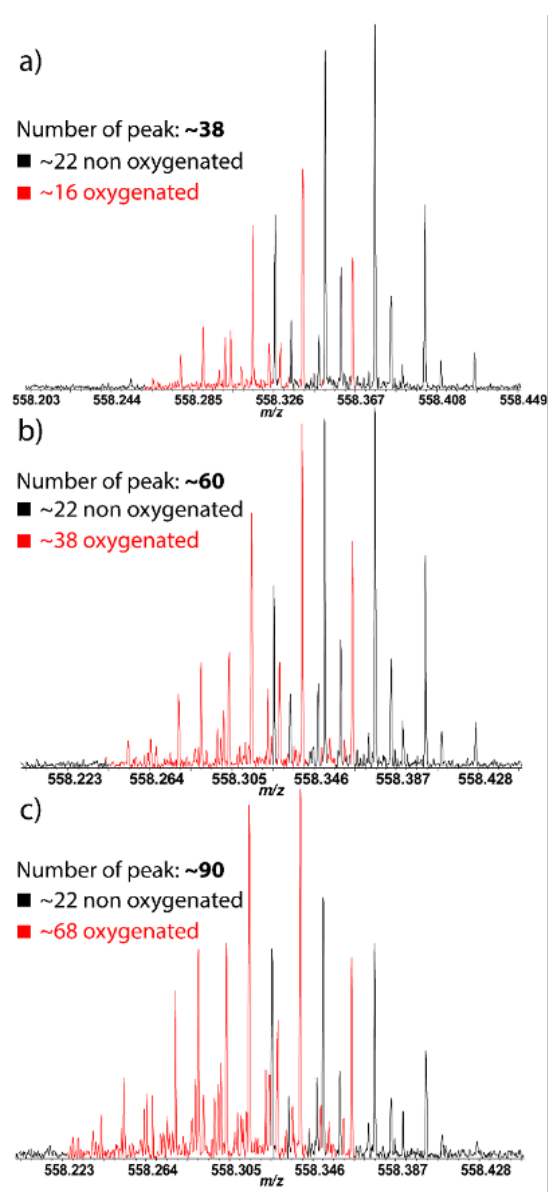

**Fig. 2:** Mass-scale zoom inset at 558.220 < m/z < 558.430 of positive-ion ESI 21T FT-ICR mass spectra for methanol extracts after a) 1-month, b) 1.8 years and c) 3.5 years.

An increase in the number of oxygen-containing compounds within the same mass range occurs from ~16 after 1 month to more than 68 after 3.5 years. Figure 2 also highlights the mass spectral peaks that contain oxygen (red) and those without oxygen (black), with the number of oxygen-containing compounds annotated. The number of peaks that do not contain oxygen in their raw formula highlights the immense complexity of molecules composing the initial sample that lead to a widely variable chemical aging reaction rate.

The number of peaks that are assigned to oxygen-containing compound classes varies from 16 in the youngest sample to 68 in the oldest sample. Thus, there are four times more oxygenated species observed after 3.5 years of aging than after 1 month, and after 1.8 years (**Fig. 2b**), the number of oxygenated species doubles. In addition, the most abundant compounds after 1 month correspond to non-oxygen containing compounds that decrease in relative abundance with time as oxygen incorporation occurs with time.

### 3.1. Amount of oxygen over time

To further highlight the oxidative processes at work in the presence of humidity, we process the mass spectral data by representing them in several formats.

First, the sum of the relative abundance or signal magnitude for oxygen-containing compounds is calculated and compared to the overall sum of the relative abundances for each sample. **Figure 3a** provides a rapid visualization of trends in the abundance of oxygen compounds detected with time, with all samples measured in triplicate. The relative abundance of oxygenated species increases logarithmically with time and rapidly in the first year from ~30% after 1 month to ~60%, and then it plateaus at approximately 70% of the total relative abundance.

Then, a sorting between the detected elemental families is performed to better understand the incorporation of oxygen into the samples. Five compound classes are defined: compounds that do not contain oxygen and $O_1$-$O_4$ classes (respectively containing 1 to 4 oxygen atoms). The sum of the relative abundances of each compound class is calculated and plotted in **Figure 3b**. As observed in **Figure 3a**, the $O_0$ families follow a logarithmic decrease. However, the other families reveal different trends. The $O_1$ family grows during the first year and then decreases while the $O_2$, $O_3$, and $O_4$ families seem to increase as a function of time. This observation suggests that species oxidize over time to incorporate more oxygen into their structure. In addition, this illustrates that the chemical aging process is complex and often proceeds through multiple generations of oxidation before the products reach greater stability that limits further oxidation.

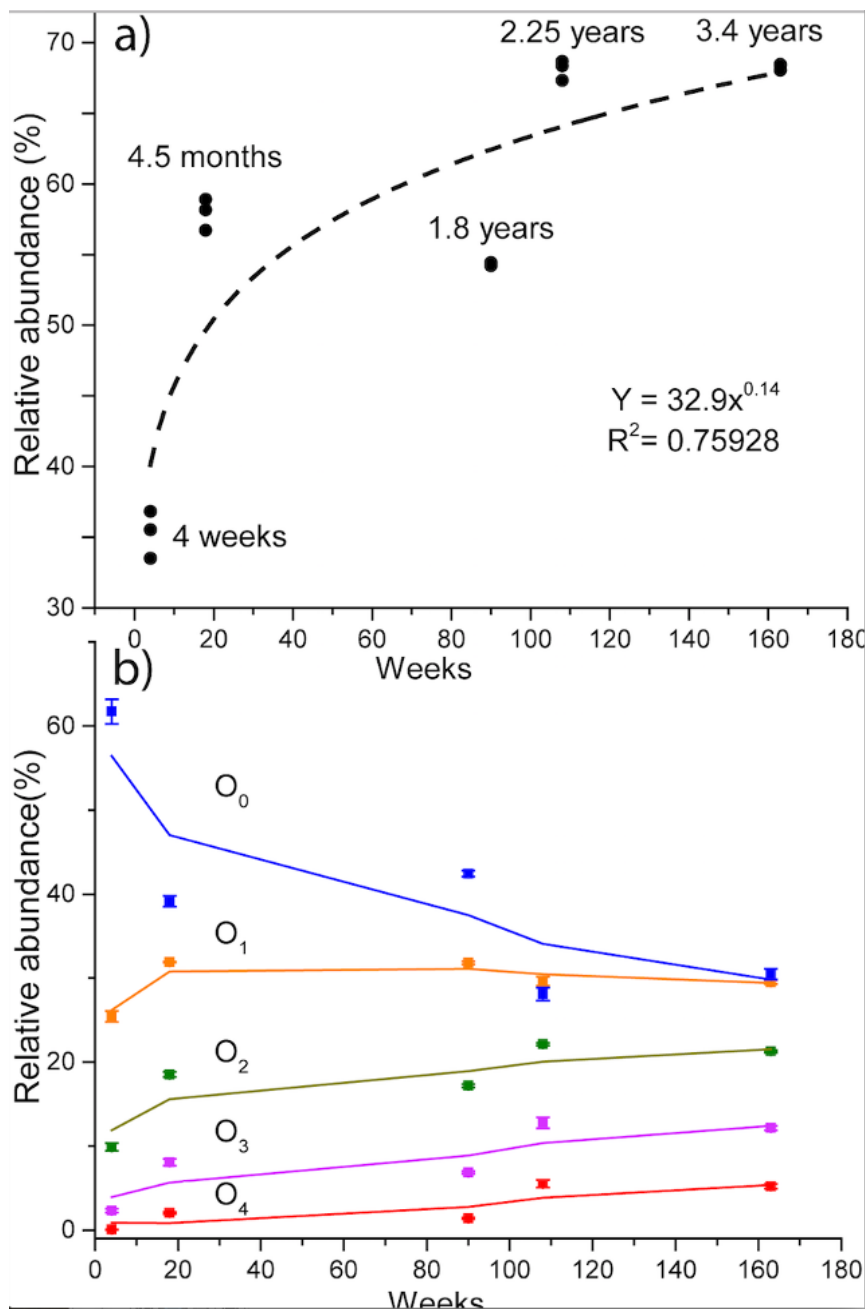

**Fig. 3:** Representation of the relative abundance of oxygenated species over time. A) Global observation of all oxygenated species. B) Relative abundance of the $O_0, O_1, O_2, O_3, O_4$ families.

### 3.2. Graphical analysis

To confirm the results obtained and facilitate the interpretation, graphical treatments are carried out. We use van Krevelen diagrams (Van Krevelen 1950). This type of graph was originally developed to assess the origin and maturity of kerogen and petroleum through the use of the H/C and O/C ratio (Kim *et al.* 2003; Wu *et al.* 2004). It is very suitable for the study of oxidation with humidity. **Figure 4a, 4b, and 4c** show the van Krevelen diagrams for the three samples aged for 1 month, 1.8 years, and 3.5 years, respectively. Here, each point is a single species present in the mass spectrum, and all species are represented. First of all, we observe that many points are located on the y-axis. Those correspond to O-free species. The other species that spread in both dimensions are oxygenated species. By comparing the three diagrams, we observe that the number of points is much larger in the third sample than in the first one. This confirms the increase in the number of oxygenated species over time observed in Figure 2. Oxygenated species are more intense as a function of time if we compare the three graphs. The H/C ratio changes very slowly, with a progressive spread of the distribution towards larger H/C ratios.

**Figures 4d, 4e,** and **4f** show the reconstructed mass spectra of the family of species that contain 8 nitrogen atoms. This family is chosen as an example because it is the most abundant one in the sample. It is divided into 5 sub-families according to the number of oxygen atoms present. Thus, we observe that the O-free family (Blue) is the major one in the first sample, but is no longer in the oldest one (3.5 years). In addition, we observe an increase in oxygenated families following the trend observed in Figure 3. This confirms the previous results concerning the amount of oxygen incorporated in the sample over time.

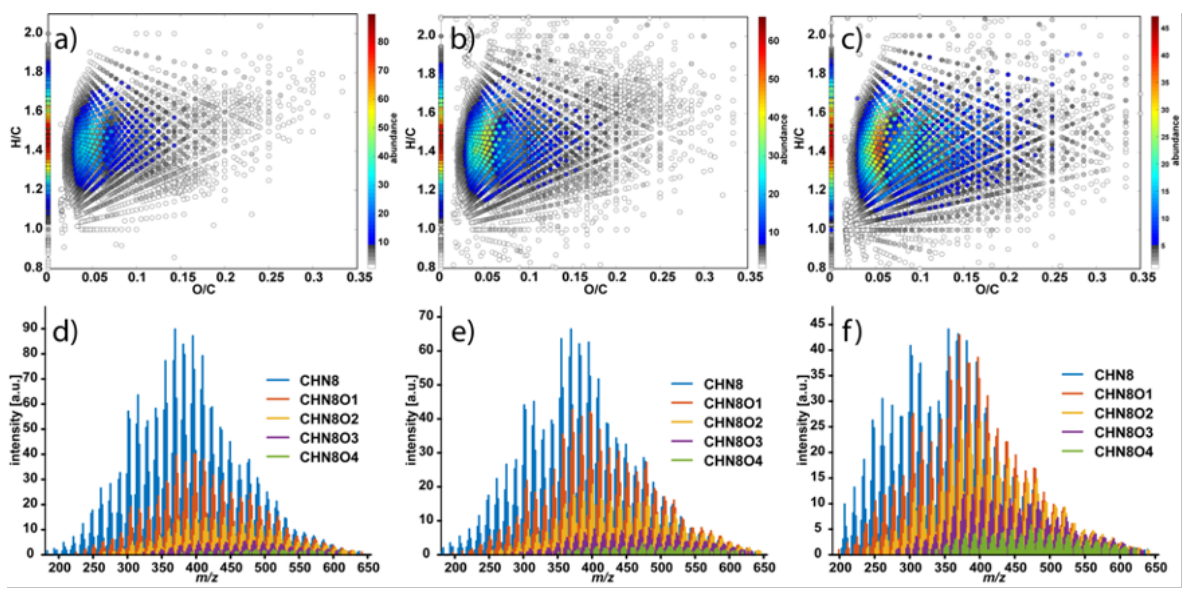

**Fig. 4.** Van Krevelen diagrams of a) 2 month old sample, b) 1.8 year old sample, c) 3.5 year old sample. The intensity of the points is given by the color code ranging from blue (lowest abundance) to red (highest abundance). Mass spectra of five different families present in the sample are given for the d) 2 month old sample, e) 1.8 year old sample, f) 3.5 year old sample.

### 3.3. Elemental analysis and quantification of the oxygen incorporation

Elemental analysis was carried out on two samples that were aged for 1 month and 2.2 years, respectively. Obtained percentages of each element composing one fresher sample and the 2.2 year old one are given in the **Table 1**. As observed, the amount of oxygen of the oldest sample is twice that of the earliest one. The H/C ratio also increases through time, in agreement with the spread towards larger H/C ratios observed on Figure 4. This global analysis validates quantitatively what is observed in detail for each molecule composing the material with mass spectrometry analyses.

Table 1. Elemental analysis of fresh and aged sample (number weighted percentages)

| Sample | %C | %H | %N | %O |
|---|---|---|---|---|
| 1 month | 36 | 40 | 20 | 4 |
| 2.2 years | 27 | 41 | 24 | 8 |

### 3.4. Infrared spectroscopy and oxygen-bearing functional groups

Infrared spectroscopy enables further study of the chemistry behind this evolution. The scope is to gain knowledge on the oxygen position and on the chemical **groups** formed. Transmission infrared spectroscopy was performed on different aged analogue samples. The aging process ranges from 1 month to 3.5 years to observe differences induced by the incorporation of oxygen. To focus on the evolution of IR absorption bands, IR spectra are normalized according to the maximum peak in each wavenumber range studied. **Figure 5** focuses on a wavenumber between 1000 $cm^{-1}$ and 1900 $cm^{-1}$. The a and c panels show the stacking of infrared spectra for the samples. The b and d panels highlight the evolution of spectra by subtracting the spectrum of the freshest sample. As observed, a band at 1700 $cm^{-1}$ increased with aging. This band is attributed to the elongation of the double-bond C=O between an oxygen and a carbon, which is consistent with carbonyl (-C(R)=O), amide (-C(NH)=O), and carboxylic acid (-C(OH)=O) groups (Long 2004). Furthermore, the band of alcohol usually present between 1280 and 1420 $cm^{-1}$ is absent in the reported IR spectrum (Long 2004). Therefore, both absences involve a preferential fixation of oxygen in C=O form rather than alcohol or ether. Panels c and d show the elongation of –CH. As observed, there is a diminution of usual –CH bands at 2800 – 2900 $cm^{-1}$ and an increase of the –CH bonded to the aldehyde (-C(H)=O) band at 2600 – 2800 $cm^{-1}$. These observations highlight the presence of aldehyde functions in our samples.

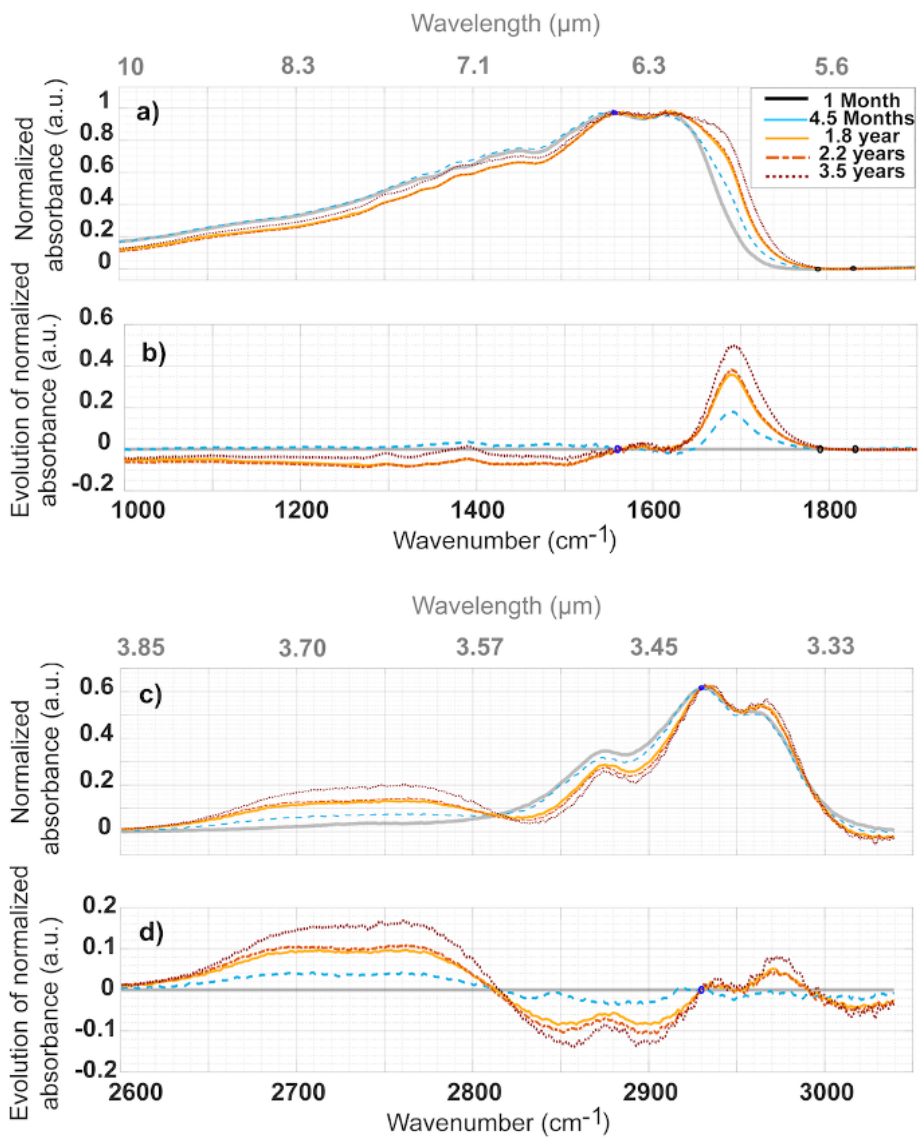

**Fig. 5:** Infrared spectra of aged analogues between 1000 cm$^{-1}$ and 1900 cm$^{-1}$. **a)** and **c)** correspond to spectra in the 1000-1800 and 2600-3000 cm$^{-1}$ range whereas **b)** and **d)** are residual spectra obtained by subtraction of the spectrum of the non-oxidized sample.

## 4. Discussion

**4.1 Mechanism at work: hydrolysis with ambient humidity**

Aerosols were exposed to humidity-controlled ambient air and may evolve through a possible contribution on the observed chemical evolution by both humidity ($H_2O$) and molecular oxygen ($O_2$). The evolution of the H/C ratios allows us to unequivocally distinguish between the two contributions. Indeed, oxidation with $O_2$ would lead to a decrease of the H/C ratio, with a change in the amount of the unsaturation, whereas hydrolysis with ambient humidity would lead to an increase of the H/C **ratio**. Therefore, the spread observed on Figure 4 towards larger H/C ratios, reinforced by the global elemental analysis reported in Table 1, clearly points to a hydrolysis process with ambient humidity. If present, the oxidation with $O_2$ has negligible effect at the time scale of the experiment. The observed evolution is the result of a hydrolysis mechanism limited by the adsorption of water vapor at the surface of the aerosol. The observation by infrared spectroscopy of the appearance and increase through time of C=O chemical bonds in the material (Figure 5) is also consistent with a progressive hydrolysis of the nitrile (-C≡N) and imine (-C=N-) chemical bonds present in the pristine aerosols.

**4.2 Role of oxygen when present since the synthesis of the aerosols**

In the present work, we considered the case of photochemical aerosols produced in dry and oxygen-free conditions. This extreme case was chosen to artificially distinguish in the laboratory the oxidation process when aerosols are exposed to humidity from the direct oxygen incorporation process during the aerosol's synthesis. However, in the atmospheres of the early-Earth and other worlds in habitable zones, water vapor may also participate in the direct synthesis of aerosols if the production occurs in humid tropospheric conditions. To our knowledge, no previous experimental simulations mimicked the formation of aerosols in the case of a $N_2/CH_4$ based atmosphere in the presence of $H_2O$.

Yet, the contribution of oxygen in the aerosol synthesis has actually been addressed, when studying experimentally the effect of CO in the case of Titan's and Pluto's atmospheres (Hörst and Tolbert 2014; Jovanović *et al.* 2020) or $CO_2$ in the case of the early Earth atmosphere (Trainer 2013). The additional presence of CO or $CO_2$ in the initial reactive mixtures led to two main results compared to the $N_2/CH_4$ reference scenario: a decrease of the total aerosol production yield and the incorporation of oxygen in the aerosol chemical composition. Molecules with up to 2 or 3 oxygen atoms were detected, among them amino-acids and nucleotide bases. In planetary conditions, oxygen is, therefore, expected to be incorporated in the aerosols both during their formation and through time by interaction of the aerosols with water vapor.

Even if the mechanisms are not yet totally unravelled, we can still discuss the relative effects of the presence of CO or $CO_2$ during aerosol synthesis on their final chemical composition as compared to their evolution via exposure to humidity. When $CO_2$ is present in a $N_2/CH_4$ reactive gas mixture, carbon incorporated into the aerosols can come from both $CO_2$ and $CH_4$ (Hicks *et al.* 2013). These results suggest that the direct incorporation of oxygen in the aerosols during the synthesis in the presence of $CO_2$, and possibly CO, would also be coupled with commensurate incorporation of carbon from the CO or $CO_2$. Under plausible planetary conditions, the effects of both processes on the final aerosol C/O ratio would, therefore, differ and require separate consideration to predict the composition of hazes in the atmospheres of planets within habitable zone.

**4.3 Timescale and viability of the process in exoplanetary atmospheres**

The experimental conditions chosen here focus on how aerosols exposed to humidity in the tropospheres of habitable zone exoplanets evolve. It does not take into account all the possible mechanisms that aerosols encounter during their time in the troposphere once they have left the relatively dry conditions of the upper troposphere. Moreover, for this first investigation, humidity was fixed at 70%. The kinetics of how the aerosols change under other humidity values will be an important parameter to be further investigated for addressing various planetary atmospheric conditions.

Nevertheless, we can already roughly discuss the order of magnitude of the timescale of the process studied here to question its viability in exoplanetary atmospheric conditions. The atmospheric humid aging process investigated here is shown to occur at a timescale of the order of a month. After four weeks, samples already contain about 35% of oxygenated molecules. Fast reactivity is observed during the first month, which means that the hazes are very prone to chemical evolution in the presence of humidity. This timescale can be compared to the estimated residence time of aerosols in the atmospheres of exoplanets in habitable zone. The easiest estimation is provided by Earth's case, where the residence time of aerosols in the troposphere is well constrained. In Earth's troposphere, dry and wet deposition removal processes limit the residence time of the aerosols in the troposphere, similarly as it possibly would occur in the lower atmosphere of exoplanets in habitable zone. The residence time, therefore, varies according to the altitude in the atmosphere, the latitude, and the meteorological conditions (Koch *et al.* 1996; Papastefanou 2008). On Earth, it is found to range between 15 days and one month, sometimes longer. The timescale of the humid oxidation process has roughly the same order of magnitude as the residence time of aerosols in the troposphere of our habitable planet. The atmospheric evolution process studied here is, therefore, likely to impact significantly the composition of the aerosols before they reach the planetary surface.

**4.4 Impact on habitability of the exoplanet**

Concerning their structure, the production of C=O chemical bonds is highlighted. These chemical bonds are consistent with the formation of carbonyl (-C(R)=O), amide (-C(NH)=O), and carboxylic acid (-C(OH)=O) functions after hydrolysis of the nitrile (-C≡N) and imine (-C=N-) groups present in the pristine aerosols. All these chemical bonds are important for prebiotic chemistry. Carbonyl bonds are precursors of amide bonds, and amide bonds are the bases of peptidic structures and sugars. Carboxylic acids, when adjacent to an amine group (-NH-) provide amino-acid functions. As discussed previously, this chemical evolution results in both an increase of the O/C ratio and the H/C ratio. This trend is illustrated by using the graphic representation proposed in Figure 6. In these, van Krevelen, areas were added according to previous works reported in the literature (Moran *et al.* 2020; Rivas-Ubach *et al.* 2018; Ruf *et al.* 2018). As observed, each region is enriched within the oxidation process, leading to a strong intensity of oxidized nitrogen-containing polycyclic aromatic hydrocarbons (N-PAH) and an increase of the intensity of lipids and amino-sugar species. A temporal conversion trend can, therefore, be hypothesized from oxidized N-PAH to lipids and amino-sugars species. This work reveals a humid oxidation process of exoplanetary haze, modifying their chemical composition through time with impact on habitability.

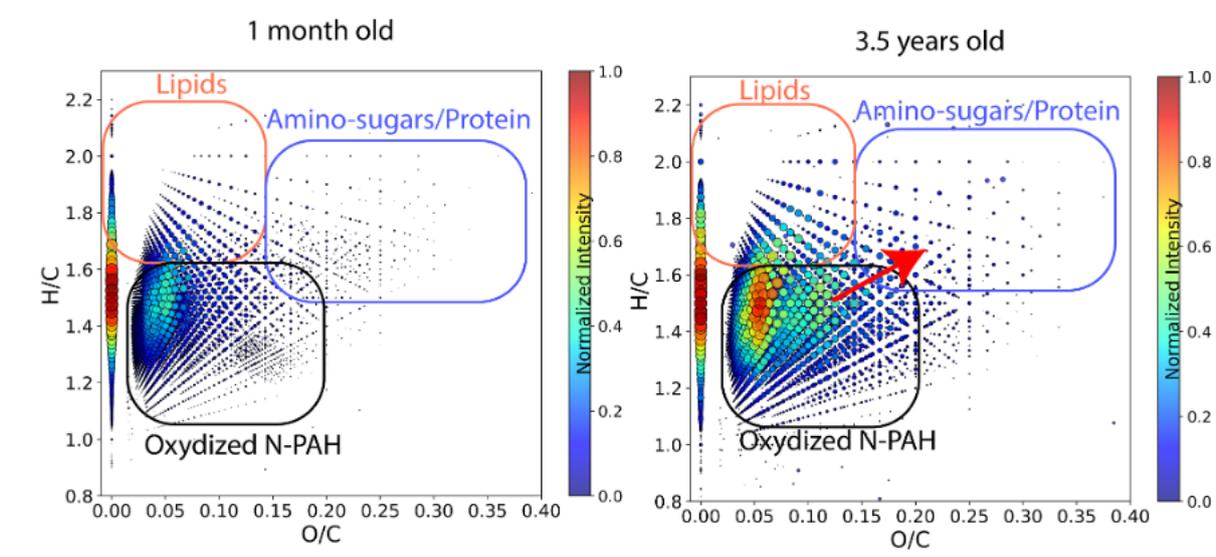

**Fig. 6:** Van Krevelen comparison between the 1 month and the 3.5 years old sample. For each graph, regions were drawn representing the area of existence of lipids, amino-sugars/proteins and oxydized N-PAH based on literature works (Rivas-Ubach *et al.* 2018; Ruf *et al.* 2018).

## 5. Conclusion

This work investigated a humid evolution process expected for organic hazes on super-Earths in habitable zone and on the early Earth during the Archean eon. O-free haze analogues were synthesized and then exposed to atmospheric humidity in controlled conditions during several years. Ultra High Resolution Mass Spectrometry with a 21 T FTICR unique instrument was used to decipher the complexity of the aerosols and to dig into the chemistry of every detected molecule. An incorporation of oxygen in their structure is observed over time. Fast reactivity occurs during the first month of aging, which means that aerosols are prone to oxidation in the presence of atmospheric humidity in agreement with the residence time of aerosols in Earth's troposphere. This humid evolution process is shown to significantly increase the carboxylic acid and carbonyl content of the haze, with impact on prebiotic chemistry. Further studies with other relative humidity conditions would interestingly broaden the description of the hydrolysis kinetics for the large range of atmospheric environments expected for extrasolar planets in their habitable zone.

## 6. Acknowledgments

N.C. thanks the European Research Council for funding via the ERC PrimChem and the ERC OxyPlanets projects (grant agreement No. 636829 and No. 101053033).

This work was supported, at COBRA Laboratory, by the European Regional Development Fund (ERDF) N° HN0001343, the European Union's Horizon 2020 Research Infrastructures program (Grant Agreement 731077), the Région Normandie, and the Laboratoire d'Excellence (LabEx) SynOrg (ANR-11-LABX-0029). Access to a CNRS FTICR research infrastructure (FR3624) is gratefully acknowledged.

Part of this work was performed at the National High Magnetic Field Laboratory, which is supported by National Science Foundation Division of Chemistry and Division of Materials Research through Cooperative Agreement No. DMR-1644779* and the State of Florida.